\documentstyle[preprint,aps]{revtex}
\draft

\def\sigz{\sigma_0}
\def\sigL{\sigma_L}
\def\LL{\Lambda_l}
\def\LM{\Lambda_m}
\def\LR{\Lambda_r}
\def\Le{\Lambda_{eff}}
\def\kL{k_l}
\def\kM{k_m}
\def\kR{k_r}
\def\yL{y_l}
\def\yR{y_r}
\def\yM{y_m}
\def\be{\begin{equation}}
\def\ee{\end{equation}}
\def\baray{\begin{eqnarray}}
\def\earay{\end{eqnarray}}

\begin{document}
\title{A Brane World Solution to the Cosmological Constant Problem}
\author{S.-H. Henry Tye\footnote
{tye@mail.lns.cornell.edu} and
Ira Wasserman\footnote
{ira@spacenet.tn.cornell.edu}}

\address{Laboratory for Nuclear Studies and Center for Radiophysics
and Space Research\\
Cornell University \\
Ithaca, NY 14853}

\medskip
\date{\today}
\maketitle


\begin{abstract}
We consider a model with two parallel (positive tension) $3$-branes 
separated by a distance $L$ in 5-dimensional spacetime.
If the interbrane space
is anti-deSitter, or is not precisely anti-deSitter but contains
no event horizons, the effective 
$4$-dimensional cosmological constant seen by observers on one of 
the branes (chosen to be the visible brane) becomes exponentially 
small as $L$ grows large.

\end{abstract}

\pacs{11.10.Kk, 04.50.+h, 98.80.-k, 98.80.Hw}

Recent observational data \cite{rev} indicate that there is 
a positive cosmological constant in the universe, which, compared to the 
Planck or the electroweak scale, is many 
orders of magnitude smaller than expected 
within the context of ordinary gravity and quantum 
field theory. This is the well-known cosmological constant problem 
\cite{weinberg}. Here we propose a solution to this problem in which 
the cosmological constant becomes exponentially small compared to 
all the other scales in the model.

The scenario is a variant of the Randall-Sundrum model\cite{RS}.
Let us consider two parallel $3$-branes (or two stacks of $3$-branes) 
in $5$-D spacetime separated by a distance $L$ in the fifth
dimension which is not necessarily
compactified. Let the brane on the left be the 
visible brane, with positive brane tension $\sigz$. The brane on 
the right has positive brane tension $\sigL\neq\sigz$, generally. Assume that 
the bulks outside the branes are anti-deSitter(AdS) spaces, with 
cosmological constants $-\LL$ and $-\LR$, respectively. 
There may be numerous solutions of the 5-D Einstein equations
for the behavior of the bulk between the branes, but let us focus
attention on the simplest, an AdS space with bulk cosmological
constant $-\LM$. (We discuss other possibilities briefly below,
and in more detail elsewhere.) We shall see that $\LM$ may be expressed
as a function of $L,\LL,\LR,\sigz,\sigL$ and the 5-D 
gravitational coupling constant $\kappa^2=8\pi G$. Thus, we regard
$\LM$ as a {\it derived} parameter; more generally, in a dynamical
5-D spacetime, $\LM$ may evolve with time as branes move
together or apart. For this simplest model for the interbrane spacetime,
we can determine the effective $4$-D 
cosmological constant $\Le$ in terms of $\LL$, $\LR$, $\sigz$, $\sigL$, 
$L$ and $\kappa^2$. 
We show that for large $L$, 
\be
\Le \approx F(\kappa,\LL,\LR,\sigz,\sigL) e^{- 2 \alpha_0 L}
\label{lamzero}
\ee
where $F$ is independent of $L$ and the positivity of the 4-D
Newton constant $G_N$ requires 
$\alpha_0= {\kappa^2 \sigz/3} - \sqrt{\kappa^2\LL/6} > 0$.
For $\alpha_0 L\gg 1$,
the effective $4$-D cosmological
constant becomes exponentially small in this
simple model without any fine tuning of the parameters.
In a $S^1/Z_2$ orbifold version of the model, the only bulk is 
between the branes, and
\be
\Le\approx 2\sigz {(\sigL+\sigz) \over(\sigL-\sigz)} e^{-\kappa^2\sigz L/3}.
\label{lamzorb}
\ee
These results remain true even if $\sigz$, which is 
simply the vacuum energy density of the $4$-D quantum field 
theory (which includes the standard model of strong and electroweak 
interactions) on the visible brane, changes due to phase transitions 
or other dynamics on the brane. 
We shall also comment briefly on solutions where the bulk between 
branes is not a pure AdS space. Demanding the absence of an event 
horizon between the branes, we again find that $\Le$ to be vanishing 
small for large $L$.
This suggests that an exponentially small effective cosmological 
constant may be a robust property of two-brane models with large
interbrane separation. One is tempted to speculate that the pure 
AdS solution is a stable fixed point as the branes move apart
and the 3-brane universe expands.

When $\alpha_0 L$ is not large, $\Le (L)$ is more complicated than
eqs.(\ref{lamzero}) and (\ref{lamzorb}).  
Furthermore, at small separations, we expect additional
nongravitational brane-brane
interactions, but
for large separations, it is reasonable to assume that the
brane dynamics is dominated by pure gravity as described here.
In a more realistic situation, the matter density on the visible 
brane (and dark matter on the other 
brane) should be included, and the separation distance $L$ should be 
treated as a dynamical variable. For slow-moving branes, a 
Born-Oppenheimer-like approach is valid, and the spacetime evolves
quasistatically from one nearly time-independent solution to another
along the sequence of two-brane models. The result resembles the 
quintessence picture \cite{track} and will be discussed elsewhere. 
Below, we derive eqs.(\ref{lamzero}) and (\ref{lamzorb}).

Consider a pair of $3$-branes at $y=0$ and $y=L$ with different positive 
brane tensions $\sigz$ and $\sigL$. Let the bulk cosmological constant be 
$-\LL$ for $y<0$, $-\LR$ for $y>L$. Choose the energy-momentum tensor 
for the bulk between the branes to have the diagonal form 
$T_{AB} = (-\lambda_0,\lambda,\lambda,\lambda,\psi)$, where $\lambda_0(y)$, 
$\lambda(y)$ and $\psi(y)$ are functions of $y$. 
Choosing $\lambda_0=\lambda$ allows us to use the metric ansatz
\be
ds^2=dy^2+A(y)[-dt^2+ e^{2Ht}\delta_{ij}dx^idx^j].
\ee
The $G_{05}$ component of the Einstein's equation 
$G_{AB}=\kappa ^2T_{AB}=8{\pi }G T_{AB}$ is satisfied trivially,
while the $G_{00}$ and the $G_{55}$ components give, respectively,
\baray
{A^{\prime\prime}\over A}&=&{2H^2\over A}+{2\kappa^2\over 3}
[\LL\Theta(-y)
+\lambda (y) \Theta(y)\Theta(L-y)\nonumber\\& &
+\LR\Theta(y-L)
-\sigz\delta(y)-\sigL\delta(y-L)]\nonumber\\
\biggl({A^\prime\over A}\biggr)^2&=&{4H^2\over A}
+{2\kappa^2\over 3}[\LL\Theta(-y)+\psi(y)\Theta(y)\Theta(L-y)\nonumber\\& &
+\LR\Theta(y-L)].
\label{einstein}
\earay
The $G_{ij}$ component does not yield additional equations (alternatively, 
with the above metric, this component requires $\lambda_0=\lambda$).
In the bulk between the branes, we have (differentiate $\psi(y)$ in eq.
[\ref{einstein}] and manipulate)
\be
\psi^\prime+{2A^\prime\over A}(\psi-\lambda)=0.
\label{psiy}
\ee
It is clear that if $\psi=\lambda$ 
then $\lambda(y)=\psi(y)=\LM$, independent of $y$.

Let us first consider this particularly simple case, in which the space
between the branes ($0<y<L$) is AdS, with a cosmological 
constant $-\LM$ that will be determined from the solution of
the $5$-D Einstein equations.
Define $\kL=\sqrt{\kappa^2\LL/6}$, $\kM=\sqrt{\kappa^2\LM/6}$ and
$\kR=\sqrt{\kappa^2\LR/6}$. 
Outside the branes, the solution for the bulks is
\baray
A(y)&=&{H^2\sinh^2[\kL(y+\yL)]\over\kL^2}\qquad(y<0)\nonumber\\
A(y)&=&{H^2\sinh^2[\kM(y+\yM)]\over\kM^2}\qquad(0<y<L)\nonumber\\
A(y)&=&{H^2\sinh^2[\kR(y+\yR)]\over\kR^2}\qquad(y>L),
\label{solution}
\earay
where $\yL$, $\yM$ and $\yR$ are constants. These bulk solutions are 
similar to those in Ref\cite{kaloper}. Continuity of the metric
at the branes implies that
\baray
{\sinh^2(\kL\yL)\over\kL^2}&=&{\sinh^2(\kM\yM)\over\kM^2}\nonumber\\
{\sinh^2[\kM(L+\yM)]\over\kM^2}&=&{\sinh^2[\kR(L+\yR)]\over\kL^2}.
\label{match}
\earay
The jump conditions at the two branes are
\baray
{\kL\cosh(\kL\yL)\over\sinh(\kL\yL)}-{\kM\cosh(\kM\yM)\over\sinh(\kM\yM)}
&=&q_0\nonumber\\
{\kM\cosh[\kM(L+\yM)]\over\sinh[\kM(L+\yM)]}
-{\kR\cosh[\kR(L+\yR)]\over\sinh[\kR(L+\yR)]}&=&q_L.
\label{jump}
\earay
where $q_0\equiv\kappa^2\sigz/3$ and $q_L\equiv\kappa^2\sigL/3$.

The expansion rate seen by observers
on the brane at $y=0$ is $H(0)=H/\sqrt{A(0)}$, where
\baray
{H^2\over A(0)}&=&{\kL^2\over\sinh^2(\kL\yL)}
={\kM^2\over\sinh^2(\kM\yM)}
\nonumber\\
&=&{[\kM^2-(\kL+q_0)^2][\kM^2-(\kL-q_0)^2]\over4q_0^2},
\label{HZ}
\earay
with $\kM^2-(\kL\pm q_0)^2>0$ or $<0$, in agreement with 
Ref\cite{horace}, which uses 
a slightly different approach. Similarly, the expansion rate seen 
by observers on the brane at $y=L$ is $H(L)=H/\sqrt{A(L)}$, 
where
\baray
{H^2\over A(L)}&=&{\kR^2\over\sinh^2[\kR(L+\yR)]}
={\kM^2\over\sinh^2[\kM(L+\yM)]}
\nonumber\\
&=&{[\kM^2-(\kR+q_L)^2][\kM^2-(\kR-q_L)^2]\over 4q_L^2},
\label{HL}
\earay
with
$\kM^2-(\kR\pm q_L)^2>0$ or $<0$. 
We can rescale $t$ so that 
$A(0)=1$, and the Hubble constants on the two branes are, respectively, 
$H(0)=H$ and $H(L)=H/\sqrt{A(L)}$.
Note that although eqs.(\ref{HZ}) and (\ref{HL}) appear to determine 
the expansion rates on the two branes completely in terms of local 
quantities (i.e., the local brane tensions, and bulk cosmological 
constants just outside each brane), the values of these quantities 
on/near the two branes are connected via $\kM$ and $\yM$.

The $4$-D Newton's constant $G_N$ can be determined by
introducing a small matter density $\rho$ to the visible brane,
that is, $q_0 \to q_0 + \kappa^2 \rho/3 $.
Requiring the Hubble constant $H$ to have the standard form
\be
H^2 \approx {{8 \pi G_N \over 3} (\Le + \rho + ...)}
\ee
yields $4 \pi q_0 G_N = \kappa^2 \alpha_0 \kL$ 
\cite{horace,cosmo1,cosmo,correct}.
Positivity of $G_N$ requires $\alpha_0=q_0-\kL>0$; 
to be specific, let us consider 
\be
0 \le \kM \le q_0 -\kL.
\label{condition}
\ee

Since the expansion rates $H$ and $H(L)$ given in eqs.(\ref{HZ}) and 
(\ref{HL}) depend on $\kM$, our goal is to express $H$ and $H(L)$ as 
functions of $L$ and the parameters $\kL$, $\kR$, $q_0$ and $q_L$. 
This requires an expression relating $L$ and $\kM$; from eqs.(\ref{match})
and (\ref{jump}) we find
\baray
\kM L =\sinh^{-1}\biggl({\kM\over H(0)}\biggr)
+\sinh^{-1}\biggl({\kM\over H(L)}\biggr) \nonumber \\
=\sinh^{-1}\biggl({2\kM q_0\over\sqrt{[\kM^2-(\kL+q_0)^2]
[\kM^2-(\kL-q_0)^2]}}\biggr)
\nonumber\\
+\sinh^{-1}\biggl({2\kM q_L\over
\sqrt{[\kM^2-(\kR+q_L)^2][\kM^2-(\kR-q_L)^2]}}\biggr) 
\label{Lfind}
\earay
Here, eq.(\ref{Lfind}) is regarded as a relation
that determines $\kM$ in terms of $L$, $q_0$, $q_L$, $\kL$ and $\kR$. 
In general, $H(0)\neq H(L)$. Because of the condition (\ref{condition}),
$H\to 0$ as $\kM \to \alpha_0=q_0-\kL$ from below. 
This means that the expansion rate as seen by observers on the visible 
brane becomes exponentially small as $L$ increases,
\be
H^2 \approx 4 \alpha_0^2 e^{-2 (\alpha_0 L - C)}
\label{Hgen}
\ee
where $\alpha_0>0$, and sinh$(C) = \kM/H(L)$.
As $\kM \to \alpha_0$ from below, $H(L)$ approaches a 
$L$-independent constant given 
by eq.(\ref{HL}) and so does $C$. 
This implies that $\Le$ becomes exponentially small as $L$ increases.

In the symmetric case, where $\kL=\kR=k$ and $q_0=q_L=q$ 
(and let $\alpha=q-k=\alpha_0$), we have
$H(L)=H(0)=H$, and $\yM = -\yL = \yR = -L/2$. The constant $C$ 
in eq.(\ref{Hgen}) becomes $C=\alpha L/2$, so  
\be
H^2 \approx 4 \alpha^2 e^{-\alpha L} + {2\kappa^2 \alpha k \over{3q}} \rho.
\ee
Thus, $\Le$ still decreases exponentially with $L$, but slower than in
the nonsymmetric case.

To this point, we have concentrated on spacetimes that are noncompact
in $y$, but similar results can be derived for the
compactified case. First, we may choose to identify $\kL=\kR=\kM$ and
derive $\kM$ and $H$ in terms of $L$. Next, we can
compactify the $y$ direction and further perform a $Z_2$
orbifold, with one brane sitting at each of the two fixed points 
($y=0,L$).
This $S^1/Z_2$ orbifold model is particularly simple, since there is 
only one bulk space between the branes sitting at the two end points.
This is an expanding (non-supersymmetric) version
of the Horava-Witten model \cite{phew}, with branes
with tension $\sigz=3q_0/\kappa^2$ at $y=0$ and
$\sigL=3q_L/\kappa^2$ at $y=L$,
separated by AdS space with bulk
cosmological constant $-6k^2/\kappa^2$. The solution
is
\be
A(y)={H^2\over k^2}\sinh^2[k(y-sy_0)]
\ee
where $s=+1$ for $y>0$ and
$s=-1$ for $y<0$.
Because of the symmetry of the model, we only
need to consider the jump conditions at $y=0$ and
$y=L$, which are $2k/q_0=\tanh ky_0$ and
$2k/q_L=\tanh[k(L-y_0)]$, respectively. 
Combining the jump conditions implies
\be
{q_0\over 2k}={\tanh kL-q_L/2k\over
1-(q_L/2k)\tanh kL};
\label{qzdet}
\ee
if $q_L/2k=\pm 1$, then $q_0/2k=\mp 1$ irrespective of $kL$, but
for $q_L/2k\neq 1$, $q_0/2k\to 1$ as $kL\to\infty$.
According to our viewpoint, eq.(\ref{qzdet}) determines 
$k$ given $q_0,q_L$ and $L$. 

The expansion rates on the branes are $H(0)=H/\sqrt{A(0)}$
and $H(L)=H/\sqrt{A(L)}$, where $H^2(0)=k^2[(q_0/2k)^2-1]$
and $H^2(L)=k^2[(q_L/2k)^2-1]$. For large values of
$q_0L$ we find that
\be
H^2(0)\approx {q_0^2(q_L+q_0)\over (q_L-q_0)}e^{-q_0L}
\label{Hz}
\ee
Thus, for $q_0L\gg 1$ and $q_L>q_0$, the cosmological
constant on the $y=0$ brane is exponentially small.
Moreover, although eq.(\ref{Hz})
may appear singular as $q_L\to q_0$, in fact 
$H^2(0)\approx q_0^2e^{-q_0L/2}$ in that case.
If $q_L=-q_0$, as considered in Ref\cite{RS},
then $y_0=\pm\infty$, so $H(0)=H(L)=0$ and therefore
$\vert q_0\vert=2k$ for {\it any} finite non-zero $L$.
Our model 
is qualitatively different, since it involves two positive 
tension branes, and interprets the bulk cosmological constant
as a parameter derived from the brane tensions and separation.

Notice that $A(y)$ given by eq.(\ref{solution}) may vanish at certain 
values of $y$. We view points where $A(y)=0$ as event horizons
since it takes a test particle leaving either brane an infinite time to 
reach them according to observers on the brane from which it is launched.
(There are no spacetime singularities at these points, since all
derivatives of $A(y)$ are finite.)
We do not worry about event horizons outside the branes ($y<0$ and $y>L$),
which recede to large distances from the branes as $L$ increases, but
an event horizon between the branes might be worrisome. To avoid an 
event horizon between branes, we need $\yM<L+\yM < 0$, which 
can be true if $\kR^2 - q_L^2 - \alpha_0^2 >0$. This condition can be
satisfied when the two branes and the outside bulks are not identical, 
but in the symmetric case,
avoiding a horizon between the branes requires
$\alpha<0$, which violates the positivity of $G_N$. So, in the 
symmetric case, we will have an event horizon right between the branes 
even when they are close together ($L \to 0$). (Non-gravitational 
interbrane interaction could alter this conclusion for very small 
$L$, but not in general.) In the orbifold case, 
although we can allow arbitrary values of $q_0$ and
$q_L$, horizons are only avoided in the model if
$L<y_0$, which implies $q_L<0$. Fortunately, horizons can
be avoided altogether if the space between the branes is not pure AdS.

There is nothing sacred about AdS space between the branes
(see \cite{nonads}), and we 
might expect non-AdS behavior to be the rule rather than the exception,
particularly at different stages in the evolution of a slowly-changing
5-D spacetime. Generally speaking, we can find infinitely
many families of solutions with different $\lambda(y)$ and $\psi(y)$
in the regions between branes.
As a first stab at a more complicated two-brane
model than we have presented here, 
we have studied the case where $\lambda=\LM$ is a 
constant, but $\psi (y)=\LM-K/A^2(y)$, which satisfies eq.(\ref{psiy})
for $y-$independent $K$.
In this case, we do not get
a relation between $\kM$ and $L$ (because $\psi(y)\neq\LM$). 
The question of whether $H^2$ is small is
related to horizon formation between the branes. If horizons
are absent, then $H^2$ must be exponentially small for large $L$.
If a horizon may only appear between branes at large $L$,
then $H^2$ must also be exponentially small.
In a dynamical solution with separating branes, we might
imagine that $H^2\to 0$ as $t\to\infty$, and no horizon forms.

To be specific, let us consider the orbifold model in some detail.
Here, we find (for $K>0$ and $A(0)\equiv 1$)
\be
A(y)=-{H^2\over 2k^2}+\biggl(1+{H^2\over 2k^2}\biggr)
{\cosh[2k(y-sy_0)]\over\cosh 2ky_0}
\ee
with $s$ chosen as for AdS.
In this case the jump conditions
at $y=0$ and $y=L$ are
\baray
{q_0\over 2k}&=&\biggl(1+{H^2\over 2k^2}\biggr)\tanh 2ky_0
\nonumber\\
{q_L\over 2k}&=&\biggl(1+{H^2\over 2k^2}\biggr)
{\sinh[2k(L-y_0)]\over A(L)\cosh 2ky_0};
\earay
using the first of these relations in the second implies
\baray
{H^2\over k}&=&{[(q_0+q_L)]\cosh 2kL-
(1+q_0q_L/2k)\sinh 2kL\over\sinh 2kL
-(q_L/2k)(1-\cosh 2kL)}\nonumber\\ 
&\approx&\biggl(q_0- 2k\biggr)
\biggl(1+{2q_Le^{-2kL}\over q_L-2k}\biggr).
\label{expanz}
\earay
Horizons are guaranteed to be absent in this model if 
\be
{H^2\over 2k^2}<{1\over\cosh 2ky_0-1}\approx 2e^{-2ky_0}.
\ee
Using eq.(\ref{expanz}) in the jump condition at $y=0$
implies $e^{-2ky_0}\approx e^{-kL}\sqrt{q_L(q_0-2k)/
q_0(q_L-2k)}$,
and so horizons are absent if 
\be
{q_0\over 2k}-1<{4q_Le^{-2kL}\over q_L-2k}
\ee
in the general case, $q_L\neq q_0$. If $q_0=q_L$, then
$e^{-2ky_0}\approx e^{-kL}$ and horizons are avoided if 
$q_0/2k-1<2e^{-kL}$.
In either case, the expansion rate is exponentially small
if there are no horizons in the solution.
 
The dynamical treatment of such a separation modulus may 
follow that discussed in Ref\cite{wise} and is under investigation.
For large separations, it is reasonable to assume that 
the brane-brane interaction is dominated by pure 
gravity as described above. At small separations, 
we expect brane-brane interactions beyond pure gravity.
In Ref\cite{cosmo} where an AdS-Schwarzschild solution is considered, 
the deviation from the pure AdS space 
red-shifts away rapidly as the universe expands. 
This leads one to conjecture that the pure AdS solution between the branes 
is a stable fixed point. If so, as branes move apart and 
the universe expands, the bulk solution remains horizon-free,
and approaches pure AdS space asymptotically.  
Clearly this issue needs careful investigation. 

One may even envision a multi-brane scenario, where the 
separation distances between branes play the roles of various 
scalar fields, as suggested by string/M theory. In particular, 
the separation of two nearby branes may play the role of an inflaton
\cite{dvali}, while the separation distance between 
two far-apart branes may play the role of
a scalar field similar to quintessence \cite{track}. Immediately
after inflation, the separation of two nearby branes is stabilized
by short-range brane-brane interactions \cite{wise,dvali}, 
when the effective cosmological constant 
is small compared to the radiation density. Then a third brane 
starts to move away while the $3$-brane universe is expanding, yielding 
an exponentially small $\Le$.

We thank Eanna Flanagan and Horace Stoica for useful discussions.
This research is partially supported by NSF (S.-H.H.T.) and NASA
(I.W.).


\begin{references}
\bibitem{correct} More precisely, $4\pi G_Nq_0=\kappa^2\alpha_0\kL
[1+2\kappa^2\Le(2\alpha_0+\kL)/3q_0\kL]$. Although $L-$dependent,
the correction is small if $G_N\Le/\kL^2\ll 1$.
\end{references}
\end{document}